\documentclass[pra,twocolumn,showpacs,superscriptaddress,floatfix, reprint]{revtex4}

\usepackage[OT1]{fontenc}
\usepackage[usenames,dvipsnames]{color}
\usepackage[latin1]{inputenc}
\usepackage[english]{babel}
\usepackage{graphicx}
\usepackage{color}
\usepackage[Gray,squaren]{SIunits}
\usepackage{xspace}
\usepackage{upgreek}
\usepackage{ulem}
\usepackage{epstopdf}
\usepackage{amssymb,amsmath,verbatim,ulem}

\newcommand{\ud}{\mathrm{d}}
\newcommand{\vbar}{{\bar{v}}}

\newcommand{\Isat}{I_{\mathrm{sat}}}

\newcommand{\eq}[1]{Eq.~\eqref{eq:#1}}

\newcommand{\eqsm}[2]{Eqs.~\eqref{eq:#1}--\eqref{eq:#2}}

\newcommand{\ket}[1]{|{#1}\rangle}
\newcommand{\bra}[1]{\langle{#1}|}
\newcommand{\erfc}{\operatorname{erfc}}

\begin{document}

\title{Linear and nonlinear magneto-optical rotation on the narrow strontium intercombination line}
\author{K. Pandey}
\affiliation{Centre for Quantum Technologies, National University of Singapore, 117543 Singapore, Singapore}
\author{C. C. Kwong}
\affiliation{School of Physical and Mathematical Sciences, Nanyang Technological University, 637371 Singapore, Singapore.}
\author{M. S. Pramod}
\affiliation{Centre for Quantum Technologies, National University of Singapore, 117543 Singapore, Singapore}
\author{D. Wilkowski}
\email[E-mail: ]{david.wilkowski@ntu.edu.sg}
\affiliation{Centre for Quantum Technologies, National University of Singapore, 117543 Singapore, Singapore}
\affiliation{School of Physical and Mathematical Sciences, Nanyang Technological University, 637371 Singapore, Singapore.}
\affiliation{MajuLab, CNRS-University of Nice-NUS-NTU International Joint Research Unit UMI 3654, Singapore}

\date{\today{}}

\begin{abstract}
   In the presence of an external static magnetic field, an atomic gas becomes optically active, showing magneto-optical rotation. In the saturated regime, the coherences among the excited substates give a nonlinear contribution to the rotation of the light polarization. In contrast with the linear magneto-optical rotation, the nonlinear counterpart is insensitive to Doppler broadening. By varying the temperature of a cold strontium gas, we observe both regimes by driving the $J=0\rightarrow J=1$ transition on the intercombination line. For this narrow transition, the sensitivity to the static magnetic field is typically three orders of magnitude larger than for a standard broad alkali transition.
\end{abstract}


\maketitle


\section{Introduction}

Since several decades, magneto-optical effect has been used for various applications, like sensitive optical magnetometry, and the study of fundamental properties in nature such as parity symmetry violation and wave transport in optically active media. For example, in a random medium, magneto-optical effect may suppress weak localization of light~\cite{PhysRevB.37.1884,lenke2000magnetic}. It also induces a weak transversal asymmetry in the radiation pattern of a Mie scatterer, leading to the appearance of a photonic Hall current~\cite{rikken1996observation}. In a periodic arrangement, strong magneto-optical effect in the microwave domain has been used to create topological edge states~\cite{wang2009observation}. All these examples show interesting similarities between electromagnetic wave transport in presence of magneto-optical effect and electronic transport in condensed matter systems.

As discovered by M. Faraday in 1845, magneto-optical effect in a nonabsorbing homogeneous medium leads to a net rotation of a linear polarization when the propagation axis is along the external magnetic field. This is known as the Faraday rotation. It naturally arises from circular birefringence, \emph{i.e.} the phase difference experienced by the two orthogonal circular polarization components of the incident linear polarization. The rotation angle of the linear polarization is given by:
\begin{equation}
\theta=V_BLB,
\end{equation}
where $B$ is the external static magnetic field, $L$ is the length of the medium, and $V_B$ is the Verdet constant of the material. For high sensitivity magnetometry, media with large Verdet constant are required. This criteria is naturally fulfilled by a medium that has sharp resonances, which show strong dispersive behavior. In this respect, extensive studies have been done on atomic vapors (see Ref.~\cite{RevModPhys.74.1153} for a review). So far, the best performances are obtained using electromagnetically induced transparency (EIT). Here, a narrow resonance emerges from the two-photon coupling of the ground state manifold which dramatically increases the sensitivity~\cite{barkov1989nonlinear,PhysRevLett.67.1855,PhysRevA.62.013808}. Importantly, the two-photon resonance is insensitive to Doppler frequency shift. This scheme, known as the nonlinear magneto-optical rotation (NMOR), has been optimized over the years using different strategies. One strategy is to increase the ground state coherence lifetime using thermal vapor cells with anti-relaxation walls~\cite{PhysRevA.62.043403} or cold gases~\cite{wojciechowski2010nonlinear}. Other strategies, including the use of squeezed light~\cite{PhysRevLett.105.053601}, optically dense media~\cite{PhysRevA.62.023810} and resonant cavity enhancement~\cite{crepaz2015cavity}, have been also implemented.

In this paper, we report the observation of magneto-optical effect on the \unit{689}{nm} $^1$S$_0\rightarrow ^3$P$_1$ intercombination line of laser cooled \textsuperscript{88}Sr atoms. Here, the transition is typically three orders of magnitude narrower than the usual alkali atoms. Thus, the sensitivity of the linear magneto-optical rotation (LMOR) is increased by the same order of magnitude. However, the ground state is non-degenerate, so NMOR due to ground state coherence in EIT scheme is absent. Instead, NMOR occurs because of the coherence in the excited substates. The interplay between LMOR and NMOR depends on both the probe intensity and the temperature of the gas. We show that NMOR is favored at high probe beam intensity, when the transition linewidth is dominated by Doppler broadening. We study as well the Voigt or Cotton-Mouton rotation, when the external magnetic field is in the plane orthogonal to the optical axis. Here, the magneto-optical rotation angle scales as a square of the magnetic field. Similar to the Faraday rotation, we observe an anomalous curvature close to zero magnetic field, which has a sign that is opposite to its linear counterpart.

The paper is organized as follows. The description of the experimental setup is given in Section~\ref{Experimental setup}. In Section~\ref{Theoretical formulation}, we derive the general expression of the LMOR in the Faraday and Voigt configurations. These two configurations under NMOR, are further analyzed in Sections~\ref{Faraday rotation} and~\ref{Voigt rotation} and compared with the experimental results. Finally, we give our conclusions in Section~\ref{Conclusions}.

\section{Experimental setup}
\label{Experimental setup}

The preparation of the laser-cooled \textsuperscript{88}Sr atoms is described in detail in Ref.~\cite{yang2015high}. After the magneto-optical trap (MOT) is switched off, the cold gas is held in a molasses for 20~ms followed by free expansion for another 20~ms, before the experiment is performed. The atomic cloud has a typical optical thickness of $b_\vbar(0)=1.0(5)$. The typical temperature of the cloud is 1.2(2)~$\mu$K.  The 4-$\sigma$ diameters of the cloud, in a plane perpendicular to the probe propagation axis, are 2.0(1)~mm and 1.0(1)~mm. By adjusting the cooling parameters, the optical thickness can be increased to 2.6 and the cloud temperature can be increased up to 15~$\mu$K.

The external static magnetic field, applied during the experiment, is controlled using a feedback loop system~\cite{kwong2015}. The system removes the other unwanted magnetic field components including the 50~Hz contribution. The magnetic field is stabilized below the 1~mG level over the duration of the experiment.

\begin{figure}
   \begin{center}
      \includegraphics[width =\linewidth]{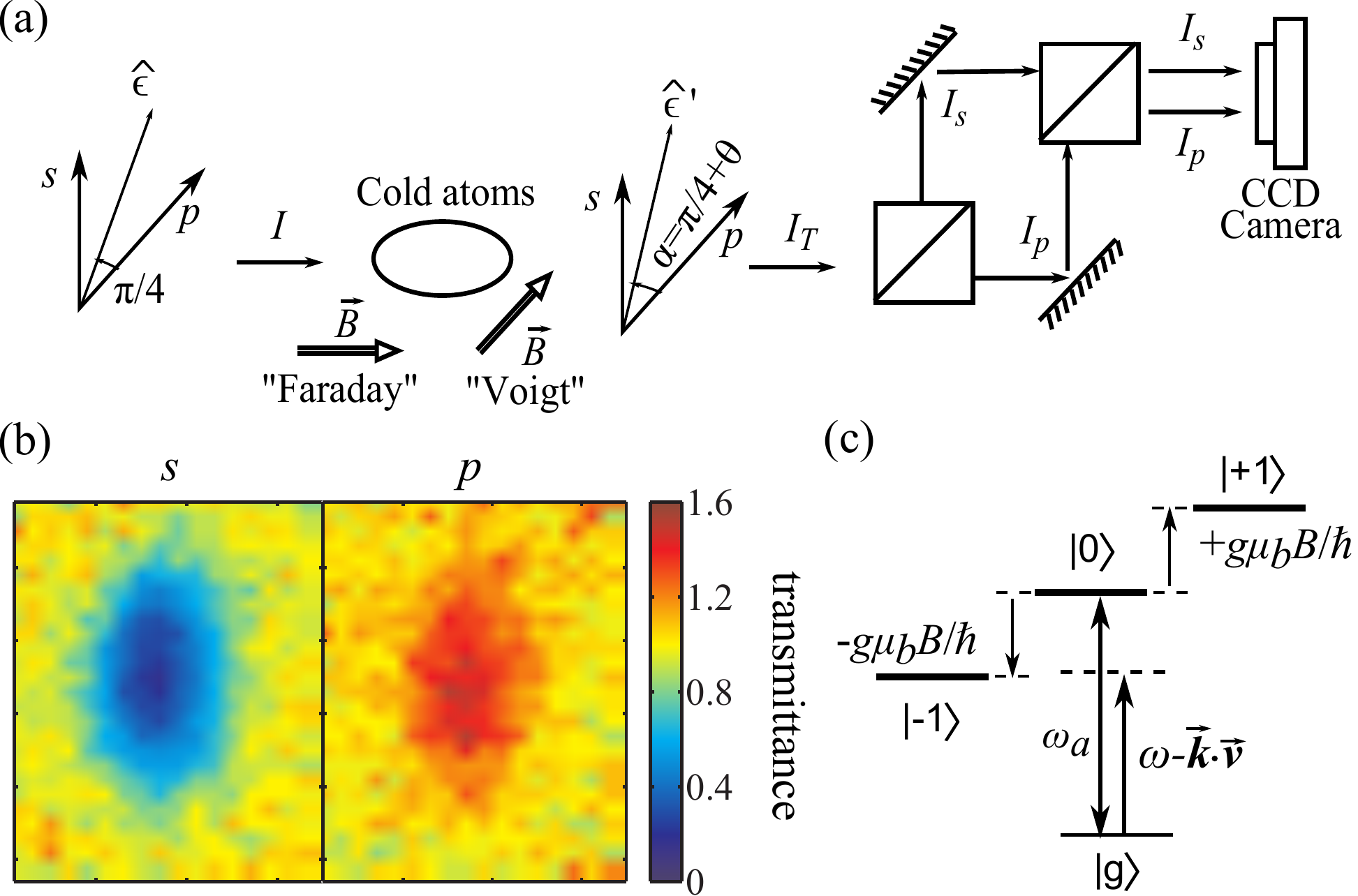}
      \caption{(a) Schematic diagram of the experimental setup. (b) Typical transmission images in the \emph{s} and \emph{p} polarization channels. (c) Energy level diagram and transition frequencies of interest in the atomic rest frame.}
      \label{fig:MORscheme}
   \end{center}
\end{figure}

The schematic diagram of the magneto-optical rotation experimental setup is depicted in Fig.~\ref{fig:MORscheme}(a). The probe is on the intercombination line resonance. Its polarization is at an angle of $45 ^\circ$ with respect to the \emph{s}-axis and the \emph{p}-axis of a polarizing cube, located after the cold atomic cloud. We define the \emph{z}-axis to be along the external static magnetic field, \emph{i.e.} $\textbf{B}=B\hat{\textbf{z}}$. Thus, in the Faraday configuration, one has $\hat{\textbf{x}}=\hat{\textbf{p}}$ and $\hat{\textbf{y}}=\hat{\textbf{s}}$. The Voigt configuration is defined such that $\hat{\textbf{z}}=\hat{\textbf{p}}$, with the propagation of the probe laser being along the \emph{y}-axis. Each configuration leads to a net magneto-optical rotation of the electric field polarization, which is denoted by $\theta$. The transmission profiles in the \emph{s} and \emph{p} polarization components are collected simultaneously on a sensitive CCD camera [see for example Fig.~\ref{fig:MORscheme}(b)]. We apply a two dimensional fitting procedure on the images to extract the transmission at the center of the cold cloud. The magneto-optical rotation angle, $\theta$, is given by the formula:
\begin{equation}
\frac{I_p-I_s}{I_p+I_s}=\sin^2\alpha-\cos^2\alpha=\sin{2\theta},
   \label{eq:signal}
\end{equation}
where $\alpha$ is the polarization angle with respect to the \emph{p}-axis, and $I_s$ and $I_p$ are the transmitted intensities in the \emph{s} and \emph{p} polarization channels at the center of the cold cloud.

\section {Theoretical formulation}
\label {Theoretical formulation}

The atomic gas is modeled by an effective slab with the same optical thickness as the value along the center of the cloud. The propagation of a plane wave in the slab is described by a Helmholtz equation. In the stationary regime, we have:
\begin{equation}
\frac{\partial^2\textbf{E}}{\partial u^2}+\frac{\omega^2}{c^2}\textbf{E}=-\frac{\omega^2}{\epsilon_0c^2}\textbf{P},
\label{eq:Helmholtz}
\end{equation}
where $u$ is the propagation distance and $\omega$ is the angular frequency of the light. $c$ is the speed of light and $\epsilon_0$ is the vacuum permittivity. $\textbf{E}=E\hat{\boldsymbol{\epsilon}}$ is the electric field and $\hat{\boldsymbol{\epsilon}}$ its polarization vector. The polarization of the medium is $\textbf{P}=\rho\textrm{Tr}\{\hat{\sigma}(u)\textbf{d}\}$, with $\rho$ being the atomic density, $\hat{\sigma}$ the atomic density operator and $\textbf{d}$ the atomic dipole operator.

We take into account the finite temperature $T$ of the gas, by performing an integration of the density operator over the velocity distribution along the propagation axis:
\begin{equation}
\hat{\sigma}(u)=\frac{1}{\sqrt{2\pi}\bar{v}}\int\ud v\hat{\sigma}(u,v)\textrm{exp}\left(-\frac{v^2}{2\bar{v}^2}\right).
\label{eq:Doppler}
\end{equation}
$\bar{v}=\sqrt{\frac{k_BT}{m_a}}$ is the thermal velocity, $k_B$ is the Boltzmann constant and $m_a$ is the mass of the atom. The temporal evolution of the density matrix is governed by the Lindblad equation:
\begin{multline}
\dot{\hat{\sigma}}(u,v)=\frac{1}{i\hbar}[H,\hat{\sigma}(u,v)]
\\+\frac{\Gamma}{2}\sum_{i=-1}^{+1} \left(2\pi_i^-\hat{\sigma}(u,v)\pi_i^+-\{\pi_i^+\pi_i^-,\hat{\sigma}(u,v)\}\right),
\label{eq:Liouville}
\end{multline}
where $\pi_i^-=\ket{g}\bra{i}$ and $\pi_i^+=\ket{i}\bra{g}$. The state $\ket{i}$ is the $m=i$ substate of $^3$P$_1$, and the state $\ket{g}$ is the ground state. $\Gamma= 2\pi \times 7.5$~kHz is the excited state spontaneous emission rate and $\hbar$ the reduced Planck constant. The Hamiltonian takes the following form:
\begin{equation}
H=\sum_i E_i \ket{i}\bra{i}-\textbf{d}\!\cdot\!\textbf{E}-\mu_zB.
\label{eq:Hamilton}
\end{equation}
$E_i$ is the bare energy of the state $\ket{i}$ for $i=g,-1,0,+1$. $B$ is an external static magnetic field oriented along the z-axis. $\mu_z$ is the z-component of the magnetic moment operator of the atom.

We consider solutions of \eqsm{Helmholtz}{Hamilton} for a $J=0\rightarrow J=1$ transition, corresponding to the $^1$S$_0\rightarrow ^3$P$_1$ intercombination line of $^{88}$Sr [see Fig.~\ref{fig:MORscheme}(c)]. For a general laser intensity $I$ and optical thickness $b_\vbar$ of the cold gas, \eqsm{Helmholtz}{Hamilton} do not have an analytical solution. Numerical integration is then performed in the steady state.

\subsection{Faraday configuration}
In the Faraday configuration, the probe laser is propagating along the magnetic field such that $\hat{\textbf{u}}=\textbf{k}/k=\hat{\textbf{z}}$. The eigenmodes of the field are the left and right circular polarizations.
We note that the excited state $\ket{0}$ remains uncoupled to the light field.

The linear regime occurs when the probe beam has an intensity $I \ll \Isat$, where $\Isat=\unit{3}{\mu W cm^{-2}}$ is the saturation intensity. Here, \eqsm{Helmholtz}{Hamilton} can be solved analytically. In the weak magnetic field limit, \emph{i.e.} $g_e\mu_bB\ll\hbar\Gamma$, the mismatch of the refractive indices of the two eigenmodes is given by~\cite{sigwarth2013multiple}:
\begin{multline}
\Delta n(v)=n_+(v)-n_-(v)\\=-2\frac{6\pi\rho}{k^{3}}\frac{1}{[1-2i(\delta-kv)/\Gamma]^2}\frac{g_e\mu_bB}{\hbar\Gamma},
\label{eq:index_v}
\end{multline}
for atoms with a velocity class $v$. $\delta=\omega-\omega_a$ is the frequency detuning and $\omega_a$ the bare atomic transition frequency. $\mu_b$ is Bohr magneton and $g_e=1.5$ is the Land\'{e} factor of the excited state. Performing the integration over the Gaussian velocity distribution, we find for $\delta=0$:
 \begin{equation}
   \Delta n=-2\frac{6\pi\rho}{k^{3}}\left[1-g\left(\frac{k \bar{v}}{\Gamma}\right)\right] \left(\frac{\Gamma}{2k \bar{v}}\right)^2\frac{g_e\mu_bB}{\hbar\Gamma}.
\end{equation}
Where $g(x) = {\sqrt{\pi/8}
\exp\left(1/8x^2\right)\erfc\left(1/\sqrt{8}x\right)/x}$. Thus, when the laser is at resonance, the index of refraction mismatch is real, indicating that the Faraday rotation comes from a pure birefringence effect. The rotation of the initial electric field polarization becomes:
\begin{equation}
   \theta=kL\Delta n/2=-b_0\left[1-g\left(\frac{k \bar{v}}{\Gamma}\right)\right] \left(\frac{\Gamma}{2k \bar{v}}\right)^2\frac{g_e\mu_bB}{\hbar\Gamma},
   \label{eq:theta}
\end{equation}
where $L$ is the slab thickness. $b_0=6\pi\rho L/k^2$ is the optical thickness assuming that the medium has a zero temperature. It is linked to the optical thickness at nonzero temperature by
${b_{\bar{v}}=b_0g(k\vbar/\Gamma)}$. \eq{theta} shows, in particular, the linear dependence of $\theta$ with respect to the optical thickness. Moreover, in the high temperature limit, $g(k\bar{v}/\Gamma) \simeq \sqrt{\pi/8}\Gamma/(k\bar{v})$, thus, we get:
\begin{equation}
   \theta=-b_{\bar{v}}\frac{g_e\mu_bB}{\hbar}\frac{1}{\sqrt{2\pi}k \bar{v}},\,\,\,\,(k\vbar/\Gamma\gg 1),
   \label{eq:slopeT}
\end{equation}
showing that $|\theta/b_{\bar{v}}|\propto1/\sqrt{T}$ for $k\bar{v}/\Gamma\gg 1$ and $g_e\mu_bB\ll\hbar\Gamma$.

For gases at low temperatures, the Faraday rotation angle reduces to,
\begin{equation}
\theta=-b_0\frac{g_e\mu_bB}{\hbar\Gamma},\,\,\,\,(k\vbar/\Gamma\ll 1).
 \end{equation}
Thus, for a probe beam tuned on the resonance of the $^{88}$Sr intercombination line, we find a sensitivity of:
\begin{equation}
   \left|\frac{1}{b_0}\frac{\ud\theta}{\ud B}\right|=\unit{0.28}{rad/mG}
   \label{eq:sensitivity_T=0}
\end{equation}
For a typical Strontium cold gas, one has $L/b_0\sim 200\,\mu$m. We then find a Verdet constant of: $|V_B|\sim \unit{10^{10}}{rad/Tm}$. It is more than three orders of magnitude larger than the values reported for standard alkali cold gasses~\cite{labeyrie2001large,nash2003linear,sigwarth2013multiple}.

\subsection{Voigt configuration}

In the Voigt configuration, the magnetic field is perpendicular to the probe propagation direction. It forms an angle of $45^\circ$ with respect to the polarization of the probe field. As depicted on Fig.~\ref{fig:MORscheme}(a), we choose $\hat{\textbf{u}}=\hat{\textbf{y}}$ and $\hat{\textbf{p}}=\hat{\textbf{z}}$. The eigenmodes of the field are the two linear polarization components along the \emph{z}-axis and the \emph{x}-axis.

For $I\ll \Isat$, and in the weak magnetic field limit, i.e. $g_e\mu_bB\ll\hbar\Gamma$, the index of refraction mismatch between the two eigenmodes for a velocity class $v$ is~\cite{sigwarth2013multiple}:
\begin{equation}
\begin{split}
\Delta n(v) &= n_z(v)-n_x(v)\\
&=2\frac{6\pi i\rho}{k^3}\frac{1}{\left[1-2i(\delta-kv)/\Gamma\right]^3}\left(\frac{g_e\mu_bB}{\hbar\Gamma}\right)^{\!2}.
\end{split}
\end{equation}

Performing the integration over the Gaussian velocity distribution, we find for $\delta=0$:
 \begin{multline}
\Delta n = -\frac{6\pi i \rho}{k^3}\left(\frac{g_e\mu_bB}{\hbar\Gamma}\right)^{\!2} \left(\frac{\Gamma}{2k\vbar}\right)^{\!2}\\
\times\left\{\left(\frac{\Gamma}{2k\vbar}\right)^2\left[1-g\!\left(\frac{k\vbar}{\Gamma}\right)\right]-g\!\left(\frac{k\vbar}{\Gamma}\right)\right\}.
   \label{eq:Voigt}
\end{multline}
Thus, when the laser is at resonance, the index of refraction mismatch is imaginary. The Voigt rotation comes from a pure dichroism effect and scales like the square of the magnetic field. If the optical thickness is small, the polarization of the initial electric field is rotated by:
\begin{equation}
   \theta=ikL\Delta n/2
   \label{eq:Voigt_n}
\end{equation}

\section{Faraday rotation}
\label{Faraday rotation}

\begin{figure}
   \begin{center}
      \includegraphics[width =\linewidth]{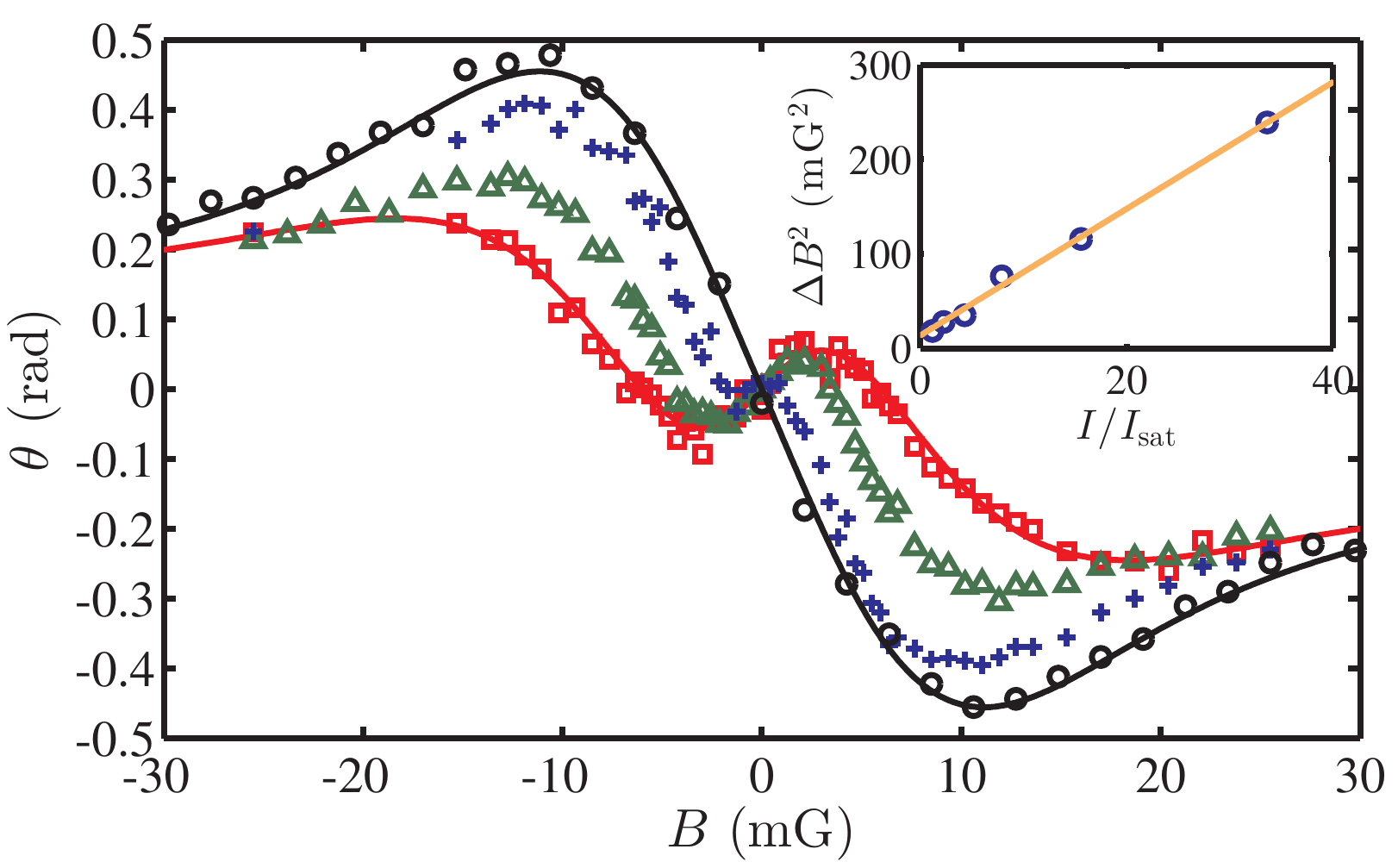}
      \caption{Rotation of the polarization of the probe laser as function of the applied magnetic field, for different values of the probe intensity. The black open circles, blue crossess, green triangles and the red squares correspond respectively to the rotation angles at $I=0.1 \Isat$, $I=2.3 \Isat$, $I=7.9 \Isat$ and $I=33.6 \Isat$. The black and red curves are results from fits (see the main text for more details). The temperature of the cold gas is $T=\unit{1}{\mu K}$. \emph{Inset:} Square of the fitted width of the NMOR signal as a function of the probe intensity. The orange line is a linear fit using the expression $\Delta B_0^2\left(1+I/(2\Isat)\right)$. We extract $\Delta B_0=3.7$ mG from the fit.}
      \label{fig:Faraday}

   \end{center}
\end{figure}

The rotation of the probe laser polarization as function of the applied magnetic field is shown in Fig.~\ref{fig:Faraday}. In the LMOR regime, \emph{i.e.} for a weak probe intensity (see the black open circles in Fig.~\ref{fig:Faraday}), we observe a dispersive-like curve with a negative slope at the origin, in agreement with \eq{theta}. The largest rotation is observed at around $|B|=$11 mG instead of $\hbar\Gamma/(g_e\mu_b)=\unit{3.6}{mG}$. This is due to the Doppler broadening associated to the cloud temperature of 1.5~$\micro$K. The profile corresponds to the Hilbert transform of a Voigt profile \cite{PhysRevA.91.032513}. We perform this transformation and fit it to the experimental data (see the black curve in Fig.~\ref{fig:Faraday}).

When the probe intensity is increased above the saturation intensity, a narrow anomalous dispersive-like curve appears with a positive slope at its origin (see Ref.~\cite{barkov1989nonlinear} for the first observation in a Samarium vapor). This anomalous Faraday rotation is associated to the NMOR effect. It emerges from the coherence $\hat{\sigma}_{-1+1}$ between the excited substates $\ket{-1}$ and $\ket{+1}$. This coherence has a lifetime of $\Gamma$, thus, it is expected to play a role if $I\gtrsim I_s$. Moreover, since $\hat{\sigma}_{-1+1}$ is due to photon redistribution between the two circular eigenmodes, it is insensitive to Doppler frequency shift. Thus, in a Doppler broadened sample, the anomalous dispersion curve could be narrower than the dispersion curve associated with LMOR. More precisely, the width of the anomalous dispersion curve is expected to be equal to the power broadened linewidth, \emph{i.e.}
\begin{equation}
\Delta B=\frac{\hbar\Gamma}{g_e\mu_b}\sqrt{1+I/(2\Isat)}.
\label{eq:Width_NMOR}
\end{equation}
Here, the intensity is divided by two, which is the intensity per eigenmode. To check this expression, we fit our data with the Hilbert transform of the LMOR Voigt profile modified with an extra Lorentzian dip at its center (see for example the red curve in Fig.~\ref{fig:Faraday}). The profile before the transform corresponds to the absorption profile of an EIT scheme (see for example Fig. 2 in ref.~\cite{PhysRevA.77.063807}). From the fit, we extract the width of the NMOR contribution which is shown in the inset in Fig.~\ref{fig:Faraday}. We plot here the square of this width to clearly indicate its linear dependency. The orange line is a linear fit using the expression $\Delta B_0^2\left(1+I/(2\Isat)\right)$. We find $\Delta B_0=3.7$ mG. Since $\hbar\Gamma/(g_e\mu_b)=\unit{3.6}{mG}$, the empirical \eq{Width_NMOR} is well verified on the experiment. We do not observe a subnatural linewidth in contrast to  some experiments done in EIT scheme~\cite{PhysRevA.77.063807,0295-5075-72-2-221}.

\begin{figure}
   \begin{center}
      \includegraphics[width =\linewidth]{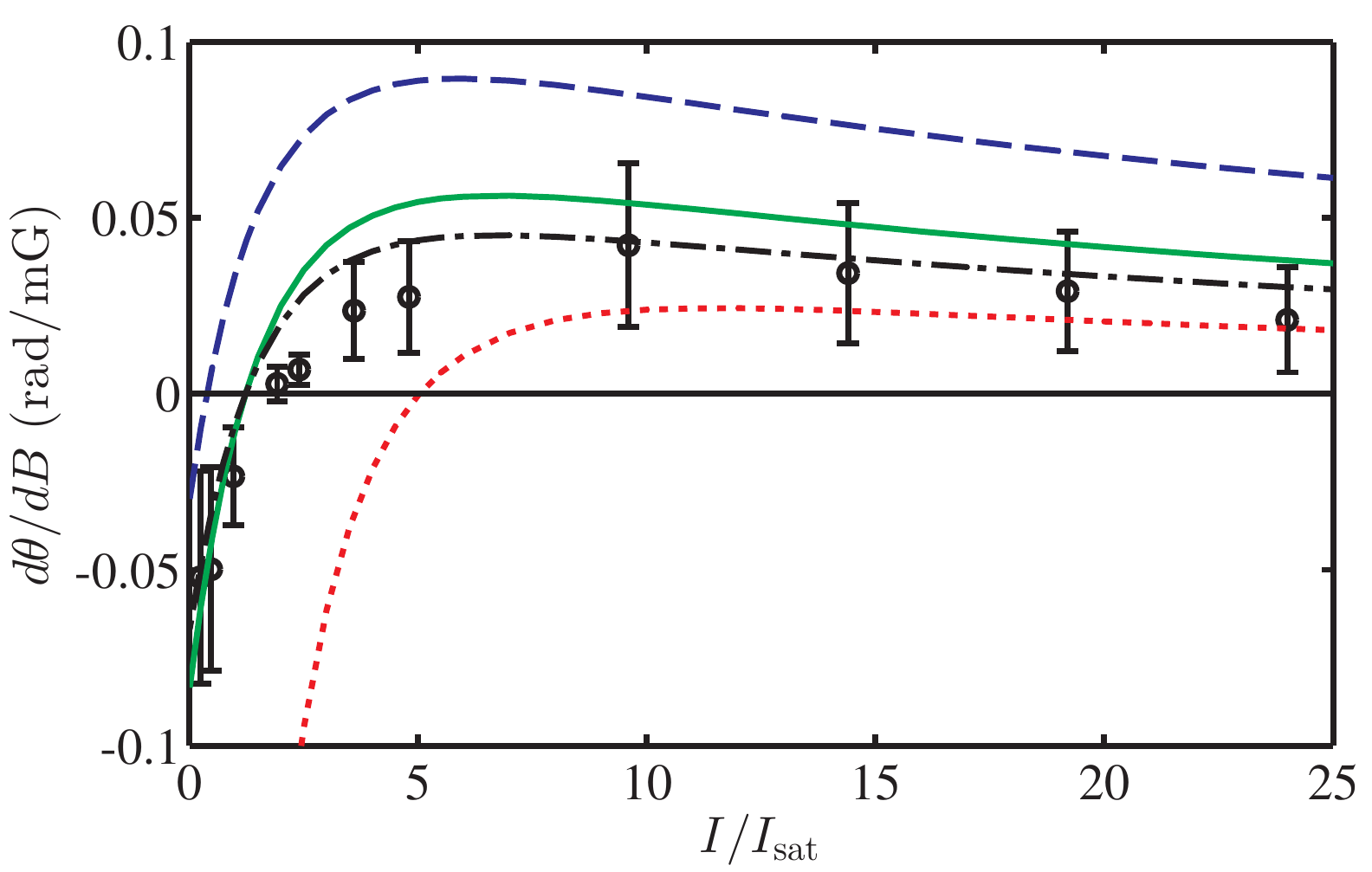}
      \caption{Sensitivity of the Faraday rotation as function of the intensity. The temperature of the cold gas is $T=1\,\mu$K, whereas its optical thickness is $b_{\bar{v}}=1.6$. Theoretical predictions are respectively at $T=0$ (red dotted curve), $T=\unit{10}{\mu K}$ (blue dashed curve), and at 1~$\mu$K, which is the same temperature as in the experiment (green solid curve). The dash-dotted curve corresponds to the theoretical curve reduced by a factor of 0.8. This is in good agreement with the experimental data (black open circles).}
      \label{fig:MOR_Int}
   \end{center}
\end{figure}

We now study the sensitivity $\ud\theta/\ud B$ of the Faraday rotation at the vicinity of $B=0$. As it is shown in Fig.~\ref{fig:Faraday}, the sensitivity is negative at low intensity and becomes positive at larger intensity. A study of the sensitivity as a function of the probe intensity, at $T=\unit{1}{\mu K}$, is summarized in Fig.~\ref{fig:MOR_Int}. The experimental sensitivity is about 0.8 times lower than the theoretical prediction corresponding to the numerical integration of \eqsm{Helmholtz}{Hamilton} (see the green solid curve in Fig.~\ref{fig:Faraday}). This discrepancy may be due to systematic effects such as finite linewidth of our probe laser~\cite{labeyrie2001large}, unbalanced polarization channel or an overestimate of the cloud optical thickness. At  $I\sim 2\Isat$, the sensitivity goes to zero, meaning that the LMOR and NMOR effects cancel out. Increasing further $I$,  the sensitivity reaches an optimal positive value at $I\sim 10\Isat$. At larger probe intensity, the sensitivity decreases smoothly toward zero because of the power broadening of the transition. We observe, in the results of the numerical integrations, the same qualitative behavior for various temperatures. However, at low temperature (see the red dashed curve for $T=0$ in Fig.~\ref{fig:MOR_Int}) the cancelation between the LMOR and NMOR effects occurs at large probe intensity. Indeed, according to \eq{theta} the LMOR is maximal at $T=0$ whereas the NMOR is temperature independent. For a similar reason, at higher temperature, the cancelation between the LMOR and NMOR effects occurs at a lower intensity of the probe beam.

\begin{figure}
   \begin{center}
\includegraphics[width =\linewidth]{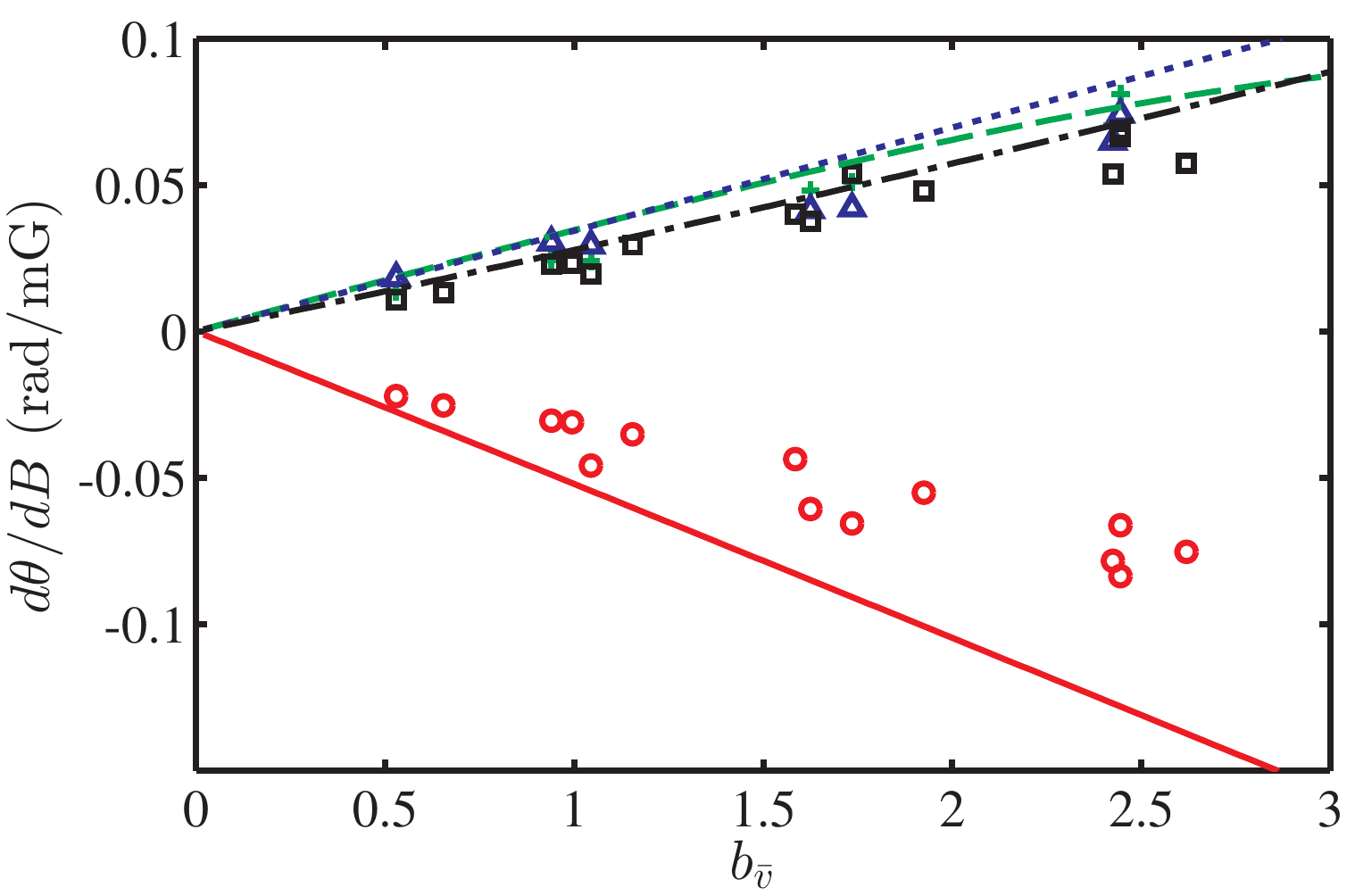}
      \caption{Sensitivity of the Faraday rotation as function of the optical thickness for different probe intensity values. The red open circles, green crosses, blue triangles and black squares correspond respectively to $I=0.1 \Isat$, $I=4.8 \Isat$, $I=8.4 \Isat$ and $I=16.8 \Isat$. The theoretical predictions for these probe intensities are, respectively, given by the red solid curve, the green dashed curve, the blue dotted curve and the black dash-dotted curve. The temperature of the cold gas is T=\unit{1}{\mu K}.}
      \label{fig:MOR_OD}
   \end{center}
\end{figure}

Next, we study the sensitivity as a function of the optical thickness. According to \eq{theta}, for a given temperature, we expect the sensitivity to be linearly dependent on the optical thickness $b_{\bar{v}}$ in the LMOR regime. It is observed in the experiment, as shown in Fig.~\ref{fig:MOR_OD} (see the red open circles). Here, the temperature is $T=\unit{1}{\mu K}$, thus the sensitivity should be reduced by a factor 0.05 with respect to the zero temperature values given in \eq{sensitivity_T=0}. The experimentally measured sensitivity are in good agreement with the theoretical prediction given by \eq{theta} even though it is slightly lower by a factor of typically 0.8. The possible reasons for this difference were pointed out earlier. In Fig.~\ref{fig:MOR_OD}, experimental data corresponding to high probe intensities are also shown. As for LMOR, we observe a linear dependency of the sensitivity with respect to the optical thickness. The optical thickness values are extracted from transmission measurements performed at low probe intensity. The numerical simulations of \eqsm{Helmholtz}{Hamilton} show that the linear dependency of the sensitivity is well fulfilled at low $b_{\bar{v}}$ but might be not correct for larger $b_{\bar{v}}$ (see for example the $I= 5\Isat$ green dashed curve at $b_{\bar{v}}\geq 2$ in Fig.~\ref{fig:MOR_OD}). The breakdown of the linear dependence on $b_\vbar$ has a simple explanation. It is due to the probe absorption in the medium. Thus, the probe saturates the medium only at its entrance but not anymore at its deeper layers. The sensitivity, therefore, becomes smaller at larger values of the optical thickness.

\begin{figure}
   \begin{center}
      \includegraphics[width =\linewidth]{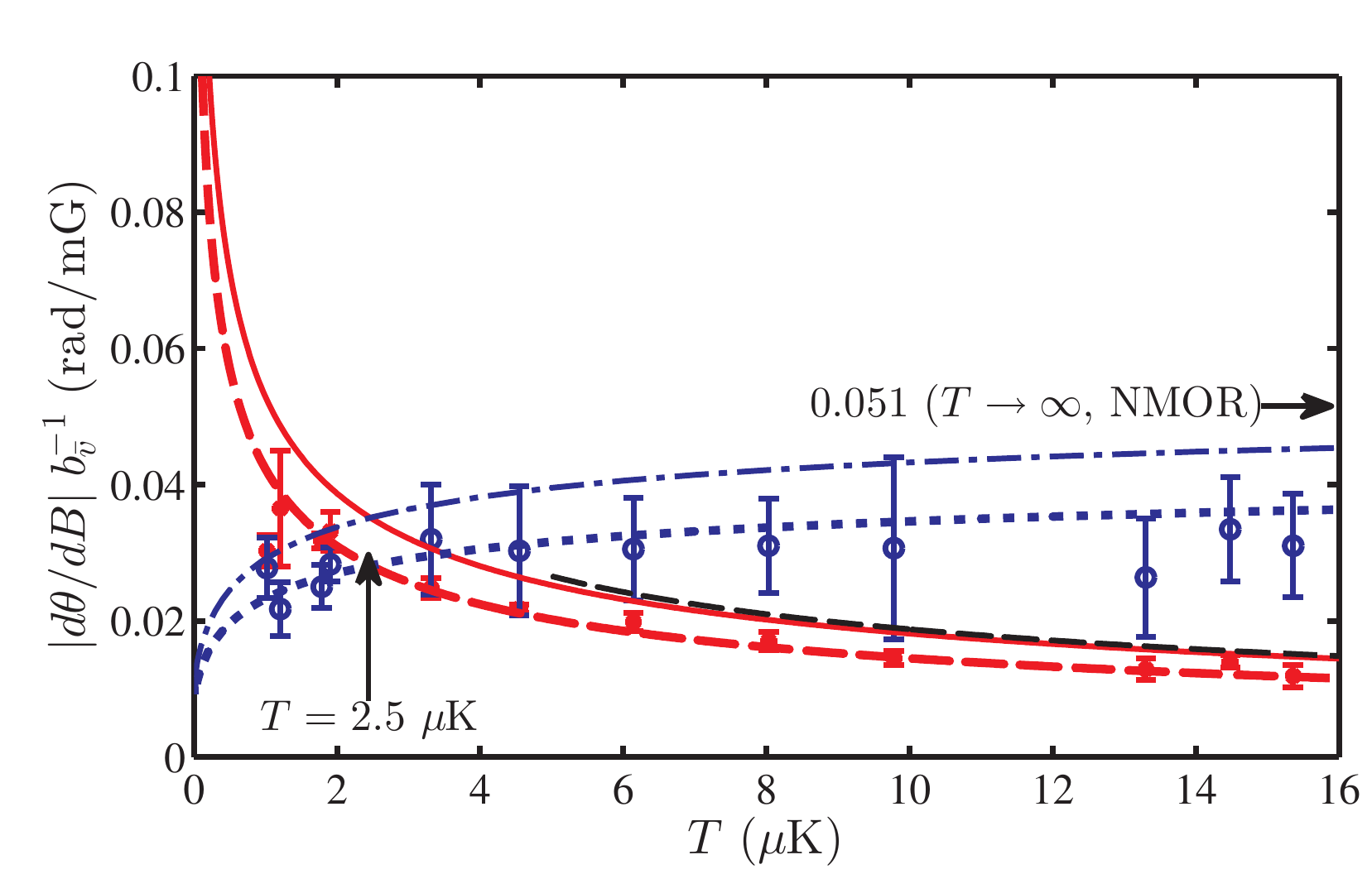}
      \caption{Absolute value of the sensitivity of the Faraday rotation as function of the temperature for intensities of $I=0.1 \Isat$ (red  dots) and $I=15 \Isat$ (blue open circles). At T=\unit{2.5}{\mu K} (vertical arrow), the sensitivity values are the same for both NMOR and LMOR. The red solid and the blue dash-dotted curves are the theoretical predictions, respectively, at low and high intensities whereas the red dashed and the short blue dotted curves are the theoretical prediction curves multiplied by 0.8. The horizontal arrow gives the predicted sensitivity in the NMOR regime when the temperature goes to infinity. The black dashed curve is the high temperature limit for LMOR corresponding to \eq{slopeT}.}
      \label{fig:MOR_Temp}
   \end{center}
\end{figure}

Finally, we discuss the dependence of the sensitivity as a function of the temperature of the cold gas. Examples in the LMOR and NMOR regimes are shown in Fig.~\ref{fig:MOR_Temp}. To avoid variations of the optical thickness due to the preparation of the cold gas at different temperature, we define a normalized sensitivity $(\ud\theta/\ud B)b_{\bar{v}}^{-1}$, with the sensitivity divided by the optical thickness. Importantly, this procedure is justified because the optical thickness is $b_{v}=0.8(0.2)$ which is weak enough for the sensitivity to be proportional to $b_{\bar{v}}$ for any probe intensity used in the experiment. In the LMOR regime, the absolute sensitivity decays as $1/\sqrt{T}$ at large temperature (see \eq{slopeT} and its representation as the black dashed curve in Fig.~\ref{fig:MOR_Temp}). The exact solution, given by \eq{theta}, is represented by the red solid curve. The red dashed curve, which is the theoretical curve scaled by a factor of 0.8, matches well with the experimental data points represented by the red dots. In the NMOR regime, the absolute sensitivity increases with temperature (see blue open circles in Fig.~\ref{fig:MOR_Temp}). Here, again the theoretical prediction has to be multiplied by a factor of 0.8 to obtain a good agreement with the experimental data points (see the dash-dotted and the dashed curves). At high temperature, the sensitivity saturates to $|\ud \theta /\ud B|b_\vbar^{-1}=\unit{0.051}{rad/mG}$. This saturation has a simple explanation. It occurs once the Doppler broadening has the same order of magnitude than the power broadening. Further increasing the temperature will not increase the number of atoms participating in the NMOR signal, leading to a saturation of the sensitivity.

\section{Voigt rotation}
\label{Voigt rotation}

\begin{figure}
   \begin{center}
      \includegraphics[width =\linewidth]{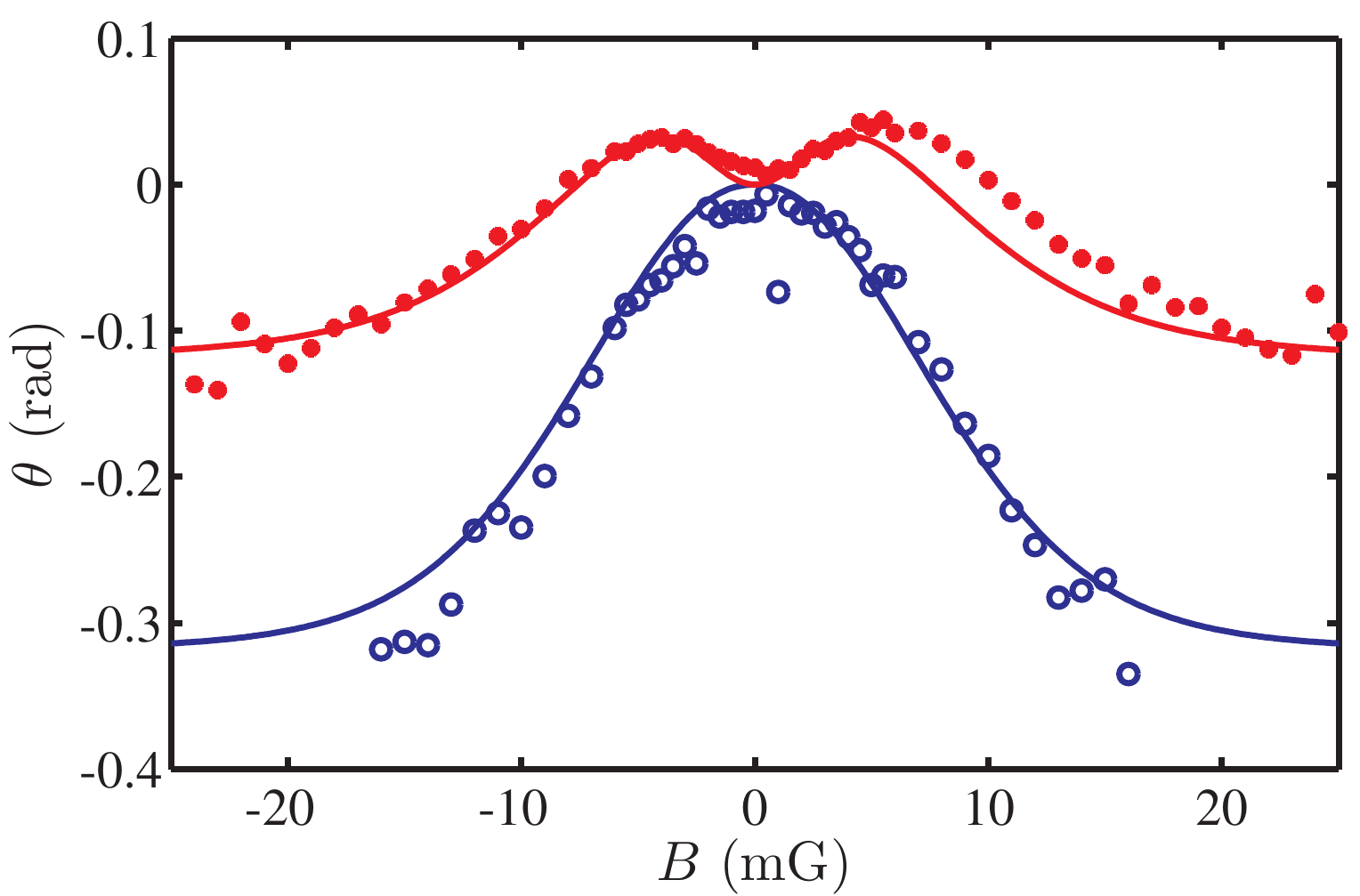}
      \caption{Magneto-optical rotation in the Voigt configuration. The data blue open circles and red filled circle correspond respectively to $I=0.1 \Isat$ and $I=12 \Isat$. The solid blue and red curves correspond to numerical simulation results. The temperature of the sample is T=\unit{1}{\mu K} and the optical thickness is $b_\vbar=1.4$.}
      \label{fig:Voigt}
   \end{center}
\end{figure}

Similar to the Faraday rotation, the Voigt rotation shows qualitatively different behavior between weak and strong probe intensity. At weak intensity (see the blue open circles in Fig.~\ref{fig:Voigt}), we observe a quadratic dependence of the rotation angle to the magnetic field, as predicted by \eqsm{Voigt}{Voigt_n}. At larger probe intensity (see the red filled circles corresponding to $I=12 \Isat$ in Fig.~\ref{fig:Voigt}), an anomalous NMOR with positive curvature is observed. This anomalous Voigt effect has a similar origin as in the case of Faraday rotation. It is due to coherence among the excited substates.  For Faraday rotation, only the $\ket{-1}$ and $\ket{+1}$ excited substates are concerned. Here, the substate $\ket{0}$ is also coupled to the light field, leading to an extra coherence contributions.

\section{Conclusions}
\label{Conclusions}

We have reported studies of the linear and nonlinear magneto-optical rotation on a cold strontium cloud using the \unit{689}{nm} intercombination line. Using moderately Doppler broadened samples, both LMOR and NMOR are clearly observed in the experiments for the Faraday and the Voigt configurations. In this regime, NMOR gives a feature narrower than LMOR because of its insensitivity to Doppler frequency shift. In particular, we show that, at large temperature, only NMOR is present and gives a significant contribution at large intensity.

To the best of our  knowledge, the strontium intercombination line is the narrowest one-photon transition used so far to study magneto-optical rotation. Our best sensitivity of $\ud\theta/\ud B=\unit{0.06}{rad/mG}$ is measured for $b_{\bar{v}}=2.6$, and T=\unit{1}{\mu K} in the linear regime. This value is roughly a factor of $50$ times smaller than the maximal value expected at $T=0$. The corresponding Verdet constant is $V_B\sim$~\unit{3\times 10^{8}}{rad/Tm}.
This value is typically two orders of magnitude larger than the values reported for standard alkali cold gas~\cite{labeyrie2001large,nash2003linear,sigwarth2013multiple}. The large Verdet constant might be promising to develop a sensitive optical magnetometer based on a low temperature strontium gas on the intercombination line. However, the LMOR regime is obtained only when the probe intensity is lower than the saturation intensity $\Isat$. Since $\Isat$ is proportional to $\Gamma$, we expect that the ultimate performance of an photons short noise limited optical magnetometer, based on LMOR, would be independent of the linewidth of the transition. In the NMOR regime such restriction on the probe intensity does not occur. Thus
a new optical magnetometer based on NMOR on a narrow transition might be considered.

\section{Acknowledgment}

This work was supported by the CQT/MoE funding grant No. R-710-002-016-271.


\begin{thebibliography}{21}
\expandafter\ifx\csname natexlab\endcsname\relax\def\natexlab#1{#1}\fi
\expandafter\ifx\csname bibnamefont\endcsname\relax
  \def\bibnamefont#1{#1}\fi
\expandafter\ifx\csname bibfnamefont\endcsname\relax
  \def\bibfnamefont#1{#1}\fi
\expandafter\ifx\csname citenamefont\endcsname\relax
  \def\citenamefont#1{#1}\fi
\expandafter\ifx\csname url\endcsname\relax
  \def\url#1{\texttt{#1}}\fi
\expandafter\ifx\csname urlprefix\endcsname\relax\def\urlprefix{URL }\fi
\providecommand{\bibinfo}[2]{#2}
\providecommand{\eprint}[2][]{\url{#2}}

\bibitem[{\citenamefont{MacKintosh and John}(1988)}]{PhysRevB.37.1884}
\bibinfo{author}{\bibfnamefont{F.~C.} \bibnamefont{MacKintosh}}
  \bibnamefont{and} \bibinfo{author}{\bibfnamefont{S.}~\bibnamefont{John}},
  \bibinfo{journal}{Phys. Rev. B} \textbf{\bibinfo{volume}{37}},
  \bibinfo{pages}{1884} (\bibinfo{year}{1988}).

\bibitem[{\citenamefont{Lenke et~al.}(2000)\citenamefont{Lenke, Lehner, and
  Maret}}]{lenke2000magnetic}
\bibinfo{author}{\bibfnamefont{R.}~\bibnamefont{Lenke}},
  \bibinfo{author}{\bibfnamefont{R.}~\bibnamefont{Lehner}}, \bibnamefont{and}
  \bibinfo{author}{\bibfnamefont{G.}~\bibnamefont{Maret}},
  \bibinfo{journal}{EPL (Europhysics Letters)} \textbf{\bibinfo{volume}{52}},
  \bibinfo{pages}{620} (\bibinfo{year}{2000}).

\bibitem[{\citenamefont{Rikken and Van~Tiggelen}(1996)}]{rikken1996observation}
\bibinfo{author}{\bibfnamefont{G.}~\bibnamefont{Rikken}} \bibnamefont{and}
  \bibinfo{author}{\bibfnamefont{B.}~\bibnamefont{Van~Tiggelen}},
  \bibinfo{journal}{Nature} \textbf{\bibinfo{volume}{381}}, \bibinfo{pages}{54}
  (\bibinfo{year}{1996}).

\bibitem[{\citenamefont{Wang et~al.}(2009)\citenamefont{Wang, Chong,
  Joannopoulos, and Solja{\v{c}}i{\'c}}}]{wang2009observation}
\bibinfo{author}{\bibfnamefont{Z.}~\bibnamefont{Wang}},
  \bibinfo{author}{\bibfnamefont{Y.}~\bibnamefont{Chong}},
  \bibinfo{author}{\bibfnamefont{J.}~\bibnamefont{Joannopoulos}},
  \bibnamefont{and}
  \bibinfo{author}{\bibfnamefont{M.}~\bibnamefont{Solja{\v{c}}i{\'c}}},
  \bibinfo{journal}{Nature} \textbf{\bibinfo{volume}{461}},
  \bibinfo{pages}{772} (\bibinfo{year}{2009}).

\bibitem[{\citenamefont{Budker et~al.}(2002)\citenamefont{Budker, Gawlik,
  Kimball, Rochester, Yashchuk, and Weis}}]{RevModPhys.74.1153}
\bibinfo{author}{\bibfnamefont{D.}~\bibnamefont{Budker}},
  \bibinfo{author}{\bibfnamefont{W.}~\bibnamefont{Gawlik}},
  \bibinfo{author}{\bibfnamefont{D.~F.} \bibnamefont{Kimball}},
  \bibinfo{author}{\bibfnamefont{S.~M.} \bibnamefont{Rochester}},
  \bibinfo{author}{\bibfnamefont{V.~V.} \bibnamefont{Yashchuk}},
  \bibnamefont{and} \bibinfo{author}{\bibfnamefont{A.}~\bibnamefont{Weis}},
  \bibinfo{journal}{Rev. Mod. Phys.} \textbf{\bibinfo{volume}{74}},
  \bibinfo{pages}{1153} (\bibinfo{year}{2002}).

\bibitem[{\citenamefont{Barkov et~al.}(1989)\citenamefont{Barkov,
  Melik-Pashayev, and Zolotorev}}]{barkov1989nonlinear}
\bibinfo{author}{\bibfnamefont{L.}~\bibnamefont{Barkov}},
  \bibinfo{author}{\bibfnamefont{D.}~\bibnamefont{Melik-Pashayev}},
  \bibnamefont{and}
  \bibinfo{author}{\bibfnamefont{M.}~\bibnamefont{Zolotorev}},
  \bibinfo{journal}{Optics communications} \textbf{\bibinfo{volume}{70}},
  \bibinfo{pages}{467} (\bibinfo{year}{1989}).

\bibitem[{\citenamefont{Scully}(1991)}]{PhysRevLett.67.1855}
\bibinfo{author}{\bibfnamefont{M.~O.} \bibnamefont{Scully}},
  \bibinfo{journal}{Phys. Rev. Lett.} \textbf{\bibinfo{volume}{67}},
  \bibinfo{pages}{1855} (\bibinfo{year}{1991}).

\bibitem[{\citenamefont{Fleischhauer et~al.}(2000)\citenamefont{Fleischhauer,
  Matsko, and Scully}}]{PhysRevA.62.013808}
\bibinfo{author}{\bibfnamefont{M.}~\bibnamefont{Fleischhauer}},
  \bibinfo{author}{\bibfnamefont{A.~B.} \bibnamefont{Matsko}},
  \bibnamefont{and} \bibinfo{author}{\bibfnamefont{M.~O.}
  \bibnamefont{Scully}}, \bibinfo{journal}{Phys. Rev. A}
  \textbf{\bibinfo{volume}{62}}, \bibinfo{pages}{013808}
  (\bibinfo{year}{2000}).

\bibitem[{\citenamefont{Budker et~al.}(2000)\citenamefont{Budker, Kimball,
  Rochester, Yashchuk, and Zolotorev}}]{PhysRevA.62.043403}
\bibinfo{author}{\bibfnamefont{D.}~\bibnamefont{Budker}},
  \bibinfo{author}{\bibfnamefont{D.~F.} \bibnamefont{Kimball}},
  \bibinfo{author}{\bibfnamefont{S.~M.} \bibnamefont{Rochester}},
  \bibinfo{author}{\bibfnamefont{V.~V.} \bibnamefont{Yashchuk}},
  \bibnamefont{and}
  \bibinfo{author}{\bibfnamefont{M.}~\bibnamefont{Zolotorev}},
  \bibinfo{journal}{Phys. Rev. A} \textbf{\bibinfo{volume}{62}},
  \bibinfo{pages}{043403} (\bibinfo{year}{2000}).

\bibitem[{\citenamefont{Wojciechowski et~al.}(2010)\citenamefont{Wojciechowski,
  Corsini, Zachorowski, and Gawlik}}]{wojciechowski2010nonlinear}
\bibinfo{author}{\bibfnamefont{A.}~\bibnamefont{Wojciechowski}},
  \bibinfo{author}{\bibfnamefont{E.}~\bibnamefont{Corsini}},
  \bibinfo{author}{\bibfnamefont{J.}~\bibnamefont{Zachorowski}},
  \bibnamefont{and} \bibinfo{author}{\bibfnamefont{W.}~\bibnamefont{Gawlik}},
  \bibinfo{journal}{Physical Review A} \textbf{\bibinfo{volume}{81}},
  \bibinfo{pages}{053420} (\bibinfo{year}{2010}).

\bibitem[{\citenamefont{Wolfgramm et~al.}(2010)\citenamefont{Wolfgramm, Cer\`e,
  Beduini, Predojevi\ifmmode~\acute{c}\else \'{c}\fi{}, Koschorreck, and
  Mitchell}}]{PhysRevLett.105.053601}
\bibinfo{author}{\bibfnamefont{F.}~\bibnamefont{Wolfgramm}},
  \bibinfo{author}{\bibfnamefont{A.}~\bibnamefont{Cer\`e}},
  \bibinfo{author}{\bibfnamefont{F.~A.} \bibnamefont{Beduini}},
  \bibinfo{author}{\bibfnamefont{A.}~\bibnamefont{Predojevi\ifmmode~\acute{c}\else
  \'{c}\fi{}}}, \bibinfo{author}{\bibfnamefont{M.}~\bibnamefont{Koschorreck}},
  \bibnamefont{and} \bibinfo{author}{\bibfnamefont{M.~W.}
  \bibnamefont{Mitchell}}, \bibinfo{journal}{Phys. Rev. Lett.}
  \textbf{\bibinfo{volume}{105}}, \bibinfo{pages}{053601}
  (\bibinfo{year}{2010}).

\bibitem[{\citenamefont{Sautenkov et~al.}(2000)\citenamefont{Sautenkov, Lukin,
  Bednar, Novikova, Mikhailov, Fleischhauer, Velichansky, Welch, and
  Scully}}]{PhysRevA.62.023810}
\bibinfo{author}{\bibfnamefont{V.~A.} \bibnamefont{Sautenkov}},
  \bibinfo{author}{\bibfnamefont{M.~D.} \bibnamefont{Lukin}},
  \bibinfo{author}{\bibfnamefont{C.~J.} \bibnamefont{Bednar}},
  \bibinfo{author}{\bibfnamefont{I.}~\bibnamefont{Novikova}},
  \bibinfo{author}{\bibfnamefont{E.}~\bibnamefont{Mikhailov}},
  \bibinfo{author}{\bibfnamefont{M.}~\bibnamefont{Fleischhauer}},
  \bibinfo{author}{\bibfnamefont{V.~L.} \bibnamefont{Velichansky}},
  \bibinfo{author}{\bibfnamefont{G.~R.} \bibnamefont{Welch}}, \bibnamefont{and}
  \bibinfo{author}{\bibfnamefont{M.~O.} \bibnamefont{Scully}},
  \bibinfo{journal}{Phys. Rev. A} \textbf{\bibinfo{volume}{62}},
  \bibinfo{pages}{023810} (\bibinfo{year}{2000}).

\bibitem[{\citenamefont{Crepaz et~al.}(2015)\citenamefont{Crepaz, Ley, and
  Dumke}}]{crepaz2015cavity}
\bibinfo{author}{\bibfnamefont{H.}~\bibnamefont{Crepaz}},
  \bibinfo{author}{\bibfnamefont{L.~Y.} \bibnamefont{Ley}}, \bibnamefont{and}
  \bibinfo{author}{\bibfnamefont{R.}~\bibnamefont{Dumke}},
  \bibinfo{journal}{Scientific reports} \textbf{\bibinfo{volume}{5}}
  (\bibinfo{year}{2015}).

\bibitem[{\citenamefont{Yang et~al.}(2015)\citenamefont{Yang, Pandey, Pramod,
  Leroux, Kwong, Hajiyev, Chia, Fang, and Wilkowski}}]{yang2015high}
\bibinfo{author}{\bibfnamefont{T.}~\bibnamefont{Yang}},
  \bibinfo{author}{\bibfnamefont{K.}~\bibnamefont{Pandey}},
  \bibinfo{author}{\bibfnamefont{M.~S.} \bibnamefont{Pramod}},
  \bibinfo{author}{\bibfnamefont{F.}~\bibnamefont{Leroux}},
  \bibinfo{author}{\bibfnamefont{C.~C.} \bibnamefont{Kwong}},
  \bibinfo{author}{\bibfnamefont{E.}~\bibnamefont{Hajiyev}},
  \bibinfo{author}{\bibfnamefont{Z.~Y.} \bibnamefont{Chia}},
  \bibinfo{author}{\bibfnamefont{B.}~\bibnamefont{Fang}}, \bibnamefont{and}
  \bibinfo{author}{\bibfnamefont{D.}~\bibnamefont{Wilkowski}},
  \bibinfo{journal}{The European Physical Journal D}
  \textbf{\bibinfo{volume}{69}}, \bibinfo{pages}{226} (\bibinfo{year}{2015}).

\bibitem[{\citenamefont{Kwong et~al.}(unpublished)\citenamefont{Kwong, Pandey,
  Pramod, Leroux, and Wilkowski}}]{kwong2015}
\bibinfo{author}{\bibfnamefont{C.~C.} \bibnamefont{Kwong}},
  \bibinfo{author}{\bibfnamefont{K.}~\bibnamefont{Pandey}},
  \bibinfo{author}{\bibfnamefont{M.~S.} \bibnamefont{Pramod}},
  \bibinfo{author}{\bibfnamefont{F.}~\bibnamefont{Leroux}}, \bibnamefont{and}
  \bibinfo{author}{\bibfnamefont{D.}~\bibnamefont{Wilkowski}}
  (\bibinfo{year}{unpublished}).

\bibitem[{\citenamefont{Sigwarth et~al.}(2013)\citenamefont{Sigwarth, Labeyrie,
  Delande, and Miniatura}}]{sigwarth2013multiple}
\bibinfo{author}{\bibfnamefont{O.}~\bibnamefont{Sigwarth}},
  \bibinfo{author}{\bibfnamefont{G.}~\bibnamefont{Labeyrie}},
  \bibinfo{author}{\bibfnamefont{D.}~\bibnamefont{Delande}}, \bibnamefont{and}
  \bibinfo{author}{\bibfnamefont{C.}~\bibnamefont{Miniatura}},
  \bibinfo{journal}{Physical Review A} \textbf{\bibinfo{volume}{88}},
  \bibinfo{pages}{033827} (\bibinfo{year}{2013}).

\bibitem[{\citenamefont{Labeyrie et~al.}(2001)\citenamefont{Labeyrie,
  Miniatura, and Kaiser}}]{labeyrie2001large}
\bibinfo{author}{\bibfnamefont{G.}~\bibnamefont{Labeyrie}},
  \bibinfo{author}{\bibfnamefont{C.}~\bibnamefont{Miniatura}},
  \bibnamefont{and} \bibinfo{author}{\bibfnamefont{R.}~\bibnamefont{Kaiser}},
  \bibinfo{journal}{Physical Review A} \textbf{\bibinfo{volume}{64}},
  \bibinfo{pages}{033402} (\bibinfo{year}{2001}).

\bibitem[{\citenamefont{Nash and Narducci}(2003)}]{nash2003linear}
\bibinfo{author}{\bibfnamefont{J.}~\bibnamefont{Nash}} \bibnamefont{and}
  \bibinfo{author}{\bibfnamefont{F.~A.} \bibnamefont{Narducci}},
  \bibinfo{journal}{Journal of Modern Optics} \textbf{\bibinfo{volume}{50}},
  \bibinfo{pages}{2667} (\bibinfo{year}{2003}).

\bibitem[{\citenamefont{Whittaker et~al.}(2015)\citenamefont{Whittaker,
  Keaveney, Hughes, and Adams}}]{PhysRevA.91.032513}
\bibinfo{author}{\bibfnamefont{K.~A.} \bibnamefont{Whittaker}},
  \bibinfo{author}{\bibfnamefont{J.}~\bibnamefont{Keaveney}},
  \bibinfo{author}{\bibfnamefont{I.~G.} \bibnamefont{Hughes}},
  \bibnamefont{and} \bibinfo{author}{\bibfnamefont{C.~S.} \bibnamefont{Adams}},
  \bibinfo{journal}{Phys. Rev. A} \textbf{\bibinfo{volume}{91}},
  \bibinfo{pages}{032513} (\bibinfo{year}{2015}).

\bibitem[{\citenamefont{Iftiquar et~al.}(2008)\citenamefont{Iftiquar, Karve,
  and Natarajan}}]{PhysRevA.77.063807}
\bibinfo{author}{\bibfnamefont{S.~M.} \bibnamefont{Iftiquar}},
  \bibinfo{author}{\bibfnamefont{G.~R.} \bibnamefont{Karve}}, \bibnamefont{and}
  \bibinfo{author}{\bibfnamefont{V.}~\bibnamefont{Natarajan}},
  \bibinfo{journal}{Phys. Rev. A} \textbf{\bibinfo{volume}{77}},
  \bibinfo{pages}{063807} (\bibinfo{year}{2008}).

\bibitem[{\citenamefont{Krishna et~al.}(2005)\citenamefont{Krishna, Pandey,
  Wasan, and Natarajan}}]{0295-5075-72-2-221}
\bibinfo{author}{\bibfnamefont{A.}~\bibnamefont{Krishna}},
  \bibinfo{author}{\bibfnamefont{K.}~\bibnamefont{Pandey}},
  \bibinfo{author}{\bibfnamefont{A.}~\bibnamefont{Wasan}}, \bibnamefont{and}
  \bibinfo{author}{\bibfnamefont{V.}~\bibnamefont{Natarajan}},
  \bibinfo{journal}{EPL (Europhysics Letters)} \textbf{\bibinfo{volume}{72}},
  \bibinfo{pages}{221} (\bibinfo{year}{2005}).

\end{thebibliography}

\end{document}